%% file: main.tex
\documentclass[acmsmall, screen]{acmart}

\AtBeginDocument{%
  }

\setcopyright{acmcopyright}
\copyrightyear{2024}
\acmYear{2024}
\acmDOI{XXXXXXX.XXXXXXX}

\acmJournal{TOSEM}

\input{setup}

\begin{document}

\newcommand{\repourl}{https://github.com/SPEAR-SE/empirical-db-issue-data}


\title{An Empirical Study on the Characteristics of Database Access Bugs in Java Applications}

\author{Wei~Liu}
\affiliation{%
  \institution{Software PErformance, Analysis, and Reliability (SPEAR) lab, Concordia University}
  \city{Montreal}
  \state{Quebec}
  \country{Canada}}
\email{w\_liu201@encs.concordia.ca}
\orcid{0000-0001-8956-730X}

\author{Shouvick~Mondal}
\affiliation{%
  \institution{Software Engineering and Testing (SET) lab, Indian Institute of Technology Gandhinagar}
  \city{Palaj}
  \state{Gujarat}
  \country{India}}
\email{shouvick.mondal@iitgn.ac.in}
\orcid{0000-0002-0703-8728}

\author{Tse-Hsun~(Peter)~Chen}
\affiliation{%
  \institution{Software PErformance, Analysis, and Reliability (SPEAR) lab, Concordia University}
  \city{Montreal}
  \state{Quebec}
  \country{Canada}}
\email{peterc@encs.concordia.ca}
\orcid{0000-0003-4027-0905}

\renewcommand{\shortauthors}{Liu et al.}

\begin{abstract}
		Database-backed applications rely on the database access code to interact with the underlying database management systems (DBMSs). 
        Although many prior studies aim at database access issues like SQL anti-patterns or SQL code smells, there is a lack of study of database access bugs during the maintenance of database-backed applications. In this paper, we empirically investigate 423 database access bugs collected from seven large-scale Java open source applications that use relational database management systems (e.g., MySQL or PostgreSQL). We study the characteristics (e.g., occurrence and root causes) of the bugs by manually examining the bug reports and commit histories. We find that the number of reported database and non-database access bugs share a similar trend but their modified files in bug fixing commits are different. Additionally, we generalize categories of the root causes of database access bugs, containing five main categories (SQL queries, Schema, API, Configuration, SQL query result) and 25 unique root causes. We find that the bugs pertaining to SQL queries, Schema, and API cover 84.2\% of database access bugs across all studied applications. In particular, SQL queries bug (54\%) and API bug (38.7\%) are the most frequent issues when using JDBC and Hibernate, respectively. Finally, we provide a discussion on the implications of our findings for developers and researchers. 

  
\end{abstract}




\maketitle


%

\input{texFiles/intro}
\input{texFiles/background}
\input{texFiles/methodology}
\input{texFiles/rq1}
\input{texFiles/rq2}

\input{texFiles/rq3}
\input{texFiles/discussion}
\input{texFiles/threats}
\input{texFiles/related}
\input{texFiles/conclusion}

\bibliographystyle{ACM-Reference-Format}
\bibliography{ref}

\end{document}

%% file: setup.tex
\usepackage{listings}
\usepackage{bm}
\usepackage{verbatim}
\usepackage{tcolorbox}
\usepackage{ragged2e}
\usepackage{amsmath}
\usepackage{stfloats}
\usepackage{textcomp}
\usepackage{booktabs, tabularx}
\usepackage{rotating}
\usepackage{xcolor}
\usepackage{soul}
\usepackage{fancybox}
\usepackage{multirow}
\usepackage{xspace}
\usepackage[ruled, lined, linesnumbered, commentsnumbered]{algorithm2e}
\usepackage{multicol,lipsum,microtype}
\usepackage{url}
\usepackage{enumerate}

\usepackage{footnote}
\usepackage{subcaption}
\usepackage{array}
\usepackage{makecell}
\usepackage{adjustbox}
\usepackage{threeparttable}

\newcommand{\REM}[1]{}
\newcommand{\etal}{{et al. }}

\definecolor{dkgreen}{rgb}{0,0.6,0}
\definecolor{gray}{rgb}{0.5,0.5,0.5}
\definecolor{mauve}{rgb}{0.58,0,0.82}
\definecolor{dkred}{rgb}{0.6, 0.1, 0}
\definecolor{dkblue}{rgb}{0, 0, 0.6}
\definecolor{lightgreen}{RGB}{205, 255, 216}
\definecolor{lightred}{RGB}{255, 220, 224}
\definecolor{lightblue}{RGB}{179, 217, 255}
\definecolor{brass}{rgb}{0.71, 0.65, 0.26}
\definecolor{violet(ryb)}{rgb}{0.53, 0.0, 0.69}
\definecolor{mediumorchid}{rgb}{0.73, 0.33, 0.83}
\definecolor{darkorange}{rgb}{1.0, 0.55, 0.0}
\definecolor{navyblue}{rgb}{0.36, 0.54, 0.66}
\definecolor{amber}{rgb}{1.0, 0.75, 0.0}

\newcommand{\tabincell}[2]{\begin{tabular}{@{}#1@{}}#2\end{tabular}}

\newcommand{\phead}[1]{\vspace{1mm} \noindent {\bf #1}}
\newcommand{\wei}[1]{\textcolor{blue}{{\it [Wei says: #1]}}}
\newcommand{\peter}[1]{\textcolor{red}{{\it [Peter says: #1]}}}

\newcommand{\shouvick}[1]{\textcolor{violet(ryb)}{{\it [Shouvick says: #1]}}}

\newcommand{\rqbox}[1]{\begin{tcolorbox}[left=4pt,right=4pt,top=4pt,bottom=4pt,colback=gray!5,colframe=gray!40!black,before skip=2pt,after skip=2pt]#1\end{tcolorbox}}

%% file: texFiles/intro.tex
\section{Introduction}\label{sec:introduction}

%
%
%
%
\noindent
From online shopping to social media, many applications need to store and access data at their back-end for rich functionalities and better user experiences. Such database-backed applications are built around the database access to interact with database management systems (DBMSs), such as MySQL, to store and retrieve data values. These database accesses are crucial for the maintenance and quality of database-backed applications.


Developers build database-backed applications to access relational databases with object-oriented programming languages such as Java, Python, C\#, PHP, and C++~\cite{MySQL_Connectors, PostgreSQL_Drivers}. Since object-oriented programming is a different paradigm compared to relational databases, developers use different technologies to ease database access by abstracting persistent data as objects. Developers often rely on two main access technologies: (i) execution of a Structured Query Language (SQL) query (e.g., using JDBC) and manually converting the results to objects; and (ii) usage of the Object Relational Mapping (ORM) frameworks, which automatically generates SQL queries and converts the result to objects based on various object-database mapping configurations. However, the inherent difference between database-backed applications and the underlying DBMS may lead to different bugs and maintenance challenges. For instance, since the syntax of SQL queries is not checked during compile time, syntax errors in SQL queries may lead to production issues. On the other hand, developers may unintentionally misuse the ORM APIs because the ORM framework hides both the generation of the underlying SQL query and its execution. As an example, when issuing calls to ORM APIs, the generated SQL queries may retrieve unused/unnecessary data from the database, thereby causing performance bugs~\cite{Chen2014_Detecting_Performance_Anti_Patterns_for_ORM}.

There are many prior works that study the maintenance issues of database-backed applications from the perspective of syntactic or semantic errors in SQL queries ~\cite{Brass_2004_Semantic_errors_in_SQL_queries, Ahadi_2016_Semantic_Mistakes_in_Writing_Seven_Different_Types_of_SQL_Queries}, SQL anti-patterns~\cite{Karwin_2010_SQL_Antipatterns_Avoiding_the_Pitfalls_of_Database_Programming, Arzamasova_2018_Cleaning_Antipatterns_in_an_SQL_Query_Log, Dintyala_2020_SQLCheck_Automated_Detection_and_Diagnosis_SQL_Anti_Patterns, Alshemaimri_2021_survey_of_problematic_database_code_fragments}, SQL code smells~\cite{The_119_SQL_Code_Smells, Sharma_2018_Measuring_and_Understanding_Database_Schema_Quality, Muse_2020_Prevalence_Impact_and_Evolution_of_SQL_Code_Smells}, and performance issues~\cite{Chen2014_Detecting_Performance_Anti_Patterns_for_ORM, Yan_2017_Understanding_Database_Performance_Inefficiencies_in_Real_World_Web_Applications, Yang_2018_Structure_Database_Backed_Web_Applications, Shao_2020_Database_Access_Performance_Antipatterns}. Specifically, \citet{Brass_2004_Semantic_errors_in_SQL_queries} proposed a list of semantic errors in SQL queries. A recent survey by~\citet{Alshemaimri_2021_survey_of_problematic_database_code_fragments} summarized categories of SQL anti-patterns and framework-specific (e.g., ORM) anti-patterns. \citet{The_119_SQL_Code_Smells} documented 119 SQL code smells while \citet{Shao_2020_Database_Access_Performance_Antipatterns} conducted a literature survey and reported 34 database access performance anti-patterns in total. Despite these efforts, there is a lack of study toward understanding database access bugs using SQL queries or ORM frameworks. Database access bugs in database-backed applications during runtime may be different from the syntactic or semantic errors in separate SQL queries since database access leverages SQL queries embedded within the application code or generated by the ORM frameworks to interact with DBMSs. On the other hand, unlike SQL anti-patterns or SQL code smells, which allow programs to execute correctly but have quality problems such as poor performance or indicate the presence of quality problems but not necessarily bugs, respectively, database access bugs may cause severe problems like crashes. Our work addresses this gap by providing categories of database access bugs from the issue tracking system and highlighting their root causes. Inspired by previous bug characterization studies~\cite{Humbatova2020_Taxonomy_Real_Faults_Deep_Learning_Systems, Shao_2020_Database_Access_Performance_Antipatterns}, we define the root cause as a human mistake in the program code, database schema, or configuration that causes database access errors.

In this paper, we conduct an empirical study to understand the characteristics and causes of database access bugs in Java database-backed applications, since Java is one of the most popular programming languages~\cite{The_2022_State_of_the_Octoverse, PYPLPopu65:online} used by millions of developers worldwide~\cite{The_2021_Java_Software_Oracle, State_of_the_Developer_Nation}. We focus on studying the systems that use relational database management systems (e.g., MySQL or PostgreSQL) due to their wide adoption and are more often used for handling complex data requests\footnote{https://www.ibm.com/cloud/blog/sql-vs-nosql}. 
We consider all types of database access bugs (e.g., not limited to performance issues that were the main focus in prior studies~\cite{Yan_2017_Understanding_Database_Performance_Inefficiencies_in_Real_World_Web_Applications, Yang_2018_Structure_Database_Backed_Web_Applications, Shao_2020_Database_Access_Performance_Antipatterns}) and consider the bugs that occur in applications that use two different types of technologies (i.e., JDBC and ORM). 
We conducted an empirical study on seven popular and large-scale open source Java database-backed applications. These applications use either the Java Database Connectivity (JDBC) or the Hibernate ORM framework for database access. 
JDBC is part of the official Java Development Kit (JDK) for accessing the DBMS and Hibernate is one of the most popular Java ORM frameworks~\cite{JAVA_Development_Research_Report}.

We collected a statistically significant sample of 5,323 fixed bug issues from the issue tracking systems of studied applications, of which 423 are manually identified as database access bugs and 4,900 are non-database access bugs. We performed a quantitative study to compare the database access bugs to non-database access bugs to study their characteristics (i.e., reported trends). Then, we performed a qualitative study on 423 database access bugs to examine their root causes. We manually examined the bug reports and commit histories of the database access bugs to understand how developers discuss/fix these bugs. The goal of the manual analysis is to identify the root cause behind the bugs and the output is categories of the root causes of database access bugs related to JDBC or Hibernate. 

In particular, we seek to answer the three following research questions (RQs):

\noindent\textbf{\hyperlink{rq1}{RQ1} (\textit{Bug occurrence}): What is the trend in the number of reported database access bugs?} 
 We find that database access bugs are reported throughout the life cycle of the applications. These observations indicate that database access bugs are common and the maintenance of database access code requires continuous attention. We also find that developers modify different sets of files in bug fixing commits for database access bugs compared to non-database access bugs, which indicates that database access bugs may have their own unique characteristics and motivates further research to examine their root causes.
	
\noindent\textbf{\hyperlink{rq2}{RQ2} (\textit{Root cause}): What are the root causes of database access bugs?} 
We generalize categories of the root causes of database access bugs. We derived the category by manually studying 423 database access bugs and identified 25 unique root causes. We find that most of the database access bugs cause problems like runtime exceptions or the return of unexpected query results to users. While a few of these bugs have been identified in the prior work \cite{Brass_2004_Semantic_errors_in_SQL_queries, Dintyala_2020_SQLCheck_Automated_Detection_and_Diagnosis_SQL_Anti_Patterns,Chen_2016_Detecting_Problems_in_Database_Access_Code_of_Large_Scale_Systems}, the majority of them have not yet been examined in the past. 

\noindent\textbf{\hyperlink{rq3}{RQ3} (\textit{Bug category}): How do categories of database access bugs prevail with different database access technologies?}
To determine whether certain categories of database access bugs arise frequently and whether this is dependent on database access technologies, we compared the percentage of categories of bugs across JDBC and Hibernate. 
Our study reveals that \textit{SQL queries}, \textit{schema}, and \textit{API} bugs cover 84.2\% database access bugs across all studied applications. Moreover, \textit{SQL queries} bug (54\%) and \textit{API} bug (38.7\%) are the most frequent issues when using JDBC and Hibernate, respectively.

The main contributions of this paper are as follows: 
\begin{itemize}
	\item To the best of our knowledge, we conduct the first empirical study of database access bugs in database-backed applications. 

	\item We find that the number of reported database and non-database access bugs share a similar trend throughout the life cycle of database-backed applications. However, their modified files in bug fixing commits are different. This implies that they are not necessarily co-located and database access bugs may have different causes and fixes as compared to non-database access bugs.

	\item We generalize categories of the root causes of database access bugs into five main categories, containing 25 unique root causes, by thoroughly studying 423 database access bugs. We find that \textit{SQL queries}, \textit{schema}, \textit{API} bugs cover 84.2\% database access bugs across all studied applications. We also find that \textit{SQL queries} bug (54\%) and \textit{API} bug (38.7\%) are the most frequent issues when using JDBC and Hibernate, respectively. We provide a discussion and implication of our findings for future work.


 
\end{itemize}

Overall, we conduct an empirical study of database access bugs in database-backed Java applications that use relational database management systems. In particular, we study the characteristics of database access bugs and the categories of their root causes. Our empirical study provides motivations and guidelines for future research to help avoid, detect, and test database access bugs in database-backed applications. Our dataset is publicly available~\cite{Replication_Package}. 

\phead{Paper organization.} The rest of this paper is organized as follows.
Section~\ref{sec:background} introduces the background of accessing databases in Java database-backed applications, using JDBC or ORM frameworks. Section~\ref{sec:methodology} describes the studied applications and our data collection approach.
Section~\ref{sec:case_study_results} presents our detailed results and Section~\ref{sec:discussion} further discusses them and provides actionable implications. Next, we discuss possible threats to validity (Section~\ref{sec:threats}) and survey related work (Section~\ref{sec:related}).
Finally, Section~\ref{sec:conclusion} concludes the paper.

%% file: texFiles/background.tex
\section{Background}\label{sec:background}

Database accesses play a central role in database-backed applications. Important business logic in such applications requires selection, insertion, or update of data in the DBMSs, such as MySQL~\cite{MySQL}. While the object in database-backed applications is often implemented in object-oriented programming languages such as Java, the database record is a row in the table defined by the database schema. Due to this discrepancy, database records need to be converted into corresponding objects in the application.


Developers often rely on two main technologies to access the underlying database and convert database records into objects in applications. The \textit{first} technology is constructing SQL queries manually and executing the SQL query by calling standard database connectivity interfaces \cite{coronel2018database} which provides an abstraction for different DBMSs. In Java, the standard database connectivity interfaces are implemented as JDBC APIs. Developers call JDBC APIs to issue the manually-constructed SQL queries to DBMSs. Subsequently, the applications retrieve the query results and developers convert them into objects. The \textit{second} technology is using Object-Relational Mapping (ORM) frameworks which provide developers a conceptual abstraction for mapping database records to objects in object-oriented languages \cite{Chen2014_Detecting_Performance_Anti_Patterns_for_ORM}. When using Hibernate (the most popular ORM framework in Java~\cite{JAVA_Development_Research_Report}), developers only need to configure the mapping between entity classes (i.e., the class mapped to a database table) and database tables, and then call Hibernate APIs. Such a mapping (configuration) allows Hibernate to automatically generate SQL queries executed by DBMSs and convert the object to/from the database record. Compared to JDBC, ORM permits developers to focus on developing the business logic without worrying too much about the database access details. When using JDBC and Hibernate, developers can choose whichever underlying DBMS they want to use. These two database access frameworks support most, if not all, relational database management systems. Developers can even switch between the underlying DBMS if needed.   

To better understand these two technologies of database access, we take Java as an example and compare the database access code using JDBC and Hibernate. Figure \ref{fig:database_access_example_JDBC} shows an example of database access using JDBC, where developers manually construct the SQL query and call the JDBC API \texttt{statement.executeQuery(query)} to issue the SQL query to the DBMS (Lines 2-3). After the SQL query is executed by the DBMS, developers retrieve the database record in query result \texttt{resultSet} and convert it into an \texttt{address} object (Lines 6-12). Figure \ref{fig:database_access_example_Hibernate} shows an example of database access using Hibernate (right) which achieves the same functionality as that of JDBC (left). The entity class \texttt{Address} (annotated with \textsf{@Entity}) is mapped to the table \texttt{address} in the DBMS using the annotation \textsf{@Table}. Developers need to set up the configuration to map each field in the \texttt{Address} entity to the database column (annotated with \textsf{@Column}) (Lines 5-10). When calling the Hibernate API \texttt{session.get(Address.class, 1)} (Line 14), Hibernate automatically generates the SQL query executed by the DBMS and serializes query results to object \texttt{address}. Then, developers can call \texttt{address.setStreet(``First  Street'')} followed by \texttt{session.update(address)} to update the corresponding database record.


\begin{figure*}
	\centering
 \begin{adjustbox}{minipage=\linewidth,scale=0.975}
	\begin{subfigure}[b]{0.47\textwidth}
		\begin{lstlisting}[frame=single, label={}, language=Java, escapechar=|, basicstyle={\tiny\ttfamily}, linewidth = {1\linewidth}, numbers=left, numbersep=5pt, xleftmargin=8pt]
// Call JDBC API to execute the SQL query
String query = "SELECT * FROM address WHERE ADDR_ID = 1"
ResultSet resultSet = statement.executeQuery(query);

// Retrieve the database record in query result ResultSet
while (resultSet.next()) {
	// Convert the database record to Java object address
	Long id = resultSet.getLong("ADDR_ID");
	String street = resultSet.getString("ADDR_STREET");
	...
	Address address = new Address(id, street, ... );
}
		\end{lstlisting}
		\caption{\texttt{JDBC}}
		\label{fig:database_access_example_JDBC}	
	\end{subfigure}
    \hfill
	\begin{subfigure}[b]{0.46\textwidth}
		\begin{lstlisting}[frame=single, label={}, language=Java, escapechar=|, basicstyle={\tiny\ttfamily}, linewidth = {1\linewidth}, numbers=left, numbersep=5pt]
// Configure the mapping
@Entity
@Table(name = "address")
public class Address {
	@Column(name = "ADDR_ID")
	private long id;
	
	@Column(name = "ADDR_STREET", length = 40 )
	private String street;
	...
}

// Call Hibernate API to fetch the database record
Address address = (Address) session.get(Address.class, 1);
		\end{lstlisting}
		\caption{\texttt{Hibernate}}
		\label{fig:database_access_example_Hibernate}
	\end{subfigure}
	
	\caption{Examples of database access using JDBC and Hibernate.}
	\label{fig:database_access_example}
 \end{adjustbox}
 \vspace{-0.4cm}
\end{figure*}


Despite the wide usage of database-backed applications, their development may be challenging due to the co-evolution of database schema and application code~\cite{Qiu_2013_Empirical_Analysis_of_Co_Evolution_of_Schema_and_Code_in_Database_Applications}. For instance, developers may be unaware of the table column name change in databases to update corresponding code in applications. Thus, accessing the database may cause runtime exceptions when the column specified in the SQL query does not exist in the database. Even worse, developers may face different challenges to access the database when using different technologies. When using JDBC, developers need to carefully construct complex SQL queries which should be free from syntax errors and be executed successfully under the database schema constraint. On the other hand, developers may unintentionally misuse the ORM APIs because ORM frameworks hide the underlying SQL query generation and execution. For instance, when calling ORM APIs, the generated SQL queries may retrieve unused/unnecessary data from the database, thereby causing performance bugs~\cite{Chen2014_Detecting_Performance_Anti_Patterns_for_ORM}. 

\REM{
Considering challenges and issues in developing database-backed applications, in this paper, we conduct an empirical study to understand the characteristics and causes of database access bugs. In particular, we want to study maintenance issues and types of bugs that occur in database access code. We address database access bugs for JDBC and Hibernate separately due to the distinct challenges developers may face when using these two technologies. We believe that our study provides motivations and guidelines for future research to resolve the challenges in developing database-backed applications by avoiding, detecting and testing database access bugs.}

In this work, we focus our attention on bugs arising from Java database access code. In particular, we set out to investigate the maintenance issues and the categories of bugs pertaining to database accesses powered by two access technologies: JDBC and ORM. Due to their distinct characteristics, the access technologies may have different underlying root causes of the associated bugs. Our empirical study aims to examine these root causes and provide actionable implications so that future development involving database-backed applications may be guided toward less bug-prone access codes. 

%% file: texFiles/methodology.tex
\section{Case Study Setup}\label{sec:methodology}

In this section, we present the setup for our empirical study. In particular, we describe our process of collecting studied applications and database access bugs.

\subsection{Collecting Studied Applications}

We focus our study on database access bugs from open-source applications implemented with Java, which is one of the most popular programming languages \cite{The_2022_State_of_the_Octoverse} used by millions of developers worldwide \cite{The_2021_Java_Software_Oracle, State_of_the_Developer_Nation}. Besides, applications in Java have been studied by much prior research \cite{Cheung_2013_Optimizing_Database_Backed_Applications, Pan2014_Guided_Test_Generation_for_Database_Applications, Chen_2016_An_Empirical_Study_on_the_Practice_of_Maintaining_Object_Relational_Mapping_Code_in_Java_System, Chen_2016_CacheOptimizer, Nagy_2018_SQLInspect, Arcuri_2019_RESTful_API_Automated_Test_Case_Generation}.

We apply three selection criteria to select the studied applications. \textit{First}, we pick the top 100 most popular Java applications that use database technology from GitHub based on the number of stars. We filter the applications using database technology by manually examining the application’s description and wiki on GitHub. \textit{Second}, a candidate application should use an issue tracking system (e.g., Jira) and contain database access bug reports so that we can study real-world bugs that occur when accessing the database. \textit{Finally}, the application should be actively maintained, having a long revision history (having more than 1,000 commits) and at least 100 fixed bug reports.

We end up with seven open-source applications in total that satisfy our selection criteria. Table \ref{tab:number_issues} shows an overview of the studied applications, such as the number of stars, lines of code (LOC), and commits. The studied applications are from various domains and four use JDBC to access the database while the other three use Hibernate. \texttt{BroadleafCommerce}~\cite{BroadleafCommerce} is an enterprise e-commerce framework while \texttt{metasfresh}~\cite{metasfresh} is an enterprise resource planning (ERP) system. \texttt{Openfire}~\cite{Openfire} is a real-time collaboration (RTC) system that supports instant messaging using the open communication protocol. \texttt{ADempiere}~\cite{ADempiere} is an enterprise business suite integrated with ERP, customer relationship management, and supply chain management. \texttt{DBeaver}~\cite{DBeaver} is a universal and multi-platform database tool which supports various popular DBMSs. 
\texttt{dotCMS}~\cite{dotCMS} is a content management system (CMS) while \texttt{OpenMRS}~\cite{OpenMRS} is a widely used patient-based electronic medical record (EMR) system.
All of the studied applications have been developed for a long period of time, ranging from 5 to 16 years. 

\begin{table*}
	\caption{The studied applications and bug issues.}
	\label{tab:number_issues}
	\centering
	\setlength{\tabcolsep}{2pt}
	\scalebox{0.7}{
	\begin{tabular}{l|l|l|r|r|r|c|r|r|r}
		\toprule
		\textbf{Application} & \textbf{Persistence}  &
		\textbf{Description} &
		\textbf{Stars(K)} &
		\textbf{LOC(K)} &
		\textbf{Commits}&
		\textbf{Study Period} &
		\tabincell{l}{\textbf{\# Fixed}\\\textbf{bug issues}}&
		\tabincell{l}{\textbf{\# Studied}\\\textbf{bug issues}}&
		\tabincell{l}{\textbf{\# Database}\\\textbf{access bugs}}\\
		\midrule
		\texttt{BroadleafCommerce}&Hibernate &E-commerce &1.5 &197 &17,381 	
		& 03/13 $\sim$ 03/17 
		& 593  	& 593 
        & 57/593 (9.6\%) \\ 
		
		\texttt{metasfresh}&JDBC&ERP System              &0.9 &1,610 &52,927 
		&06/16 $\sim$ 03/21 
		& 712  	& 712 
        & 27/712 (3.8\%) \\ 
		
		\texttt{Openfire}&JDBC& RTC Server         		&2.4 &117 &10,034 	
		&08/05 $\sim$ 04/21	
		& 749  	& 749 
		& 41/749 (5.5\%) \\
		 
		\texttt{ADempiere}&JDBC& Business Suite			&0.6 &879 &16,006 	
		& 09/15 $\sim$ 03/21 	
		& 866  	& 866 
		& 63/866 (7.3\%) \\\midrule
		
		\texttt{DBeaver}&JDBC& Database Tool            &22.5 &430 &21,176 	
		& 10/15 $\sim$ 06/21 
		& 3,203 &801 
		&106/801 (13.2\%)  \\ 
		
		\texttt{dotCMS}&Hibernate&CMS     				&0.6 &522 &18,317 	
		& 03/12 $\sim$ 05/21 
		& 4,607 &867 
		& 50/867 (5.8\%)  \\ 
		
		\texttt{OpenMRS}&Hibernate& EMR System    		& 1.0 &128 &11,530 	
		& 03/06 $\sim$ 03/21  
		& 2,359 &735 
		&79/735 (10.7\%)\\ 
		\bottomrule
	\end{tabular}}
	\footnotesize{*Note that, we study the issues of BroadleafCommerce until 03/2017 because developers do not actively maintain the issue tracking system for the open source version. However, developers are still actively developing the application.}
  \vspace{-0.2cm}
\end{table*}

\subsection{Collecting Database Access Bugs}
We collect the database access bugs from the issue tracking system of the studied applications. We first collect fixed bug issues and filter the issues using database-related keywords. Finally, we manually verify whether the identified issues are related to database access bugs. Below, we discuss our bug collection process in detail. 


\phead{Collecting studied bug issues.} 
We collect the issues with the type of \textit{bug} and fix status of \textit{fixed} and \textit{resolved} in each application during the study period, from the time they were first reported to 06/2021. Note that, we study the issues of \texttt{BroadleafCommerce} until 03/2017 because developers do not actively maintain the issue tracking system for the open source version of \texttt{BroadleafCommerce}. However, developers are still actively developing the application~\cite{BroadleafCommerce}. 
Table \ref{tab:number_issues} shows the number of fixed bug issues for each application. Then we collect studied bug issues from the fixed bug issues. For applications with fixed bug issues of less than 1,000 (i.e., the first four applications), we study all the fixed bug issues. For other applications which have 2,359 to 4,607 fixed bug issues, it is not feasible to manually study all the bug reports. Therefore, we randomly take a statistically significant sample from the fixed bug issues with a 95\% confidence level and a 3\% margin of error.


\phead{Filtering \& verifying studied bug issues.} 
We filter the studied issues by searching for database-related keywords in each issue's title, description, and comments to get the database-related issues. The database-related keywords are concluded by manually examining 50 random database-related issues from the issue tracking system and are listed below along with matching text (underlined) examples:

\begin{itemize}
	\item database: ``the entry is added in the \underline{database}''
	\item constraint: ``a foreign key \underline{constraint} fails''
	\item MySQL, PostgreSQL, SQLite, Oracle, DB2, SQL Server: ``PubSubManager: DELETE FROM ofPubsubItem LEFT JOIN breaks \underline{MySQL}''
	\item SQL: ``You have an error in your \underline{SQL} syntax''
	\item Hibernate: ``\underline{Hibernate} startup errors''
	\item JDBC: ``I have confirmed that this is an issue with the mysql \underline{JDBC} driver version.''
\end{itemize}

Next, we verify the database-related issues manually to examine the database access bug or non-database access bug because our heuristic approach (i.e., filtering by keywords) may result in false positives. 
For instance, BroadleafCommerce\#378 reports an issue that contains the keyword ``database'' in the sentence ``the data is saved appropriately in the database, but the list grid value for the column is not immediately updated''. However, according to the sentence, the issue is not a database access bug since the program accesses the database correctly. Instead, the issue is caused by calling an API incorrectly which leads to incorrect UI (i.e., list grid) refresh.

After the filtering and verifying process, we end up with manually verified database access bugs for each studied application, shown in the last column of Table \ref{tab:number_issues}. Other bugs in the studied bug issues are identified as non-database access bugs. The percentages of database access bugs among studied bug issues are from 3.8\% to 13.2\%. In total, we collected 423 database access bugs, among which 186 are related to Hibernate and 237 are related to JDBC. We also collected 4,900 non-database access bugs. 

%% file: texFiles/rq1.tex
\section{Case study results}
\label{sec:case_study_results}
In this section, we present our case study results by answering the research questions (RQs). 

\hypertarget{rq1}{}
\subsection{RQ1: What is the Trend in the Number of Reported Database Access Bugs?}

\label{rq:rq1} 

\phead{Motivation.}
In this research question, we study when database access bugs are reported in the studied applications. In particular, we want to examine whether database access bugs are more likely to occur at a specific time period. For example, are most database access bugs reported at the beginning of the development history and fewer bugs are reported as the application evolves, or are the bugs reported throughout the development history? We are also interested in the trend of the number of database access bugs reported, because database access is critical to database-backed applications and any failures related to database access may have disastrous results~\cite{Cost_of_Database_Downtime}. Knowing when database access bugs are reported is the first step toward better debugging and maintenance of database-backed applications.
\phead{Approach.}
We study how many and when the database access bugs collected in Section~\ref{sec:methodology} are reported across the study period. For each studied application, we study the number of reported bugs. For each bug, we take the creation time of the bug report as the reporting time. 
To gain a more comprehensive understanding of the quantitative magnitude of database access bugs, we further study the correlation between the reported numbers of database access bugs and non-database access bugs throughout the study period (i.e., every three or six months).
We choose to use Spearman's rank correlation $ r_{s} $ because it is a non-parametric correlation test and does not have an assumption on the underlying data distribution~\cite{zwillinger_crc_2000}. We classify the strength of the relationship between the number of database access bugs and non-database access bugs to \textit{zero} (0), \textit{weak} ($ \pm $ 0.1 to $ \pm $0.3), \textit{moderate} ($ \pm $ 0.4 to $ \pm $ 0.6), \textit{strong} ($ \pm $ 0.7 to $ \pm $ 0.9), and \textit{perfect} ($ \pm $ 1) according to the $ r_{s} $ value threshold as per prior work~\cite{dancey2007statistics}.

\begin{figure*}
	\centering
	\begin{subfigure}[b]{0.32\textwidth}
		\centering
		\includegraphics[width=\textwidth]{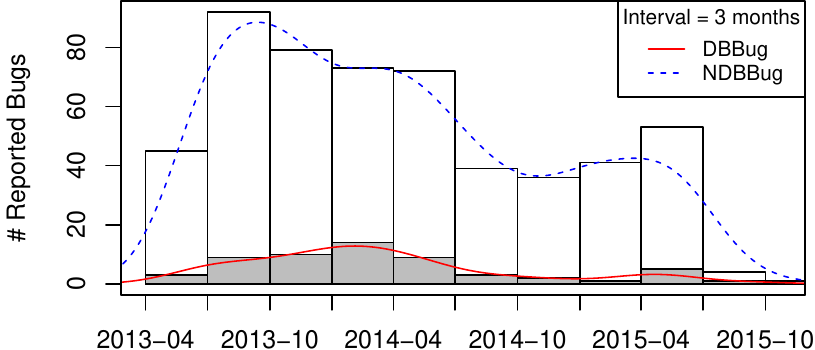}	
		\caption{\texttt{BroadleafCommerce}}
	\end{subfigure}
	\begin{subfigure}[b]{0.33\textwidth}
		\includegraphics[width=\textwidth]{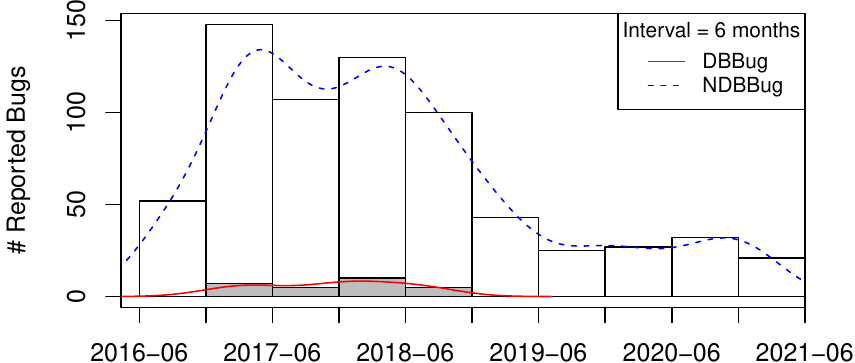}
		\caption{\texttt{metasfresh}}
	\end{subfigure}
	\begin{subfigure}[b]{0.32\linewidth}
		\centering
		\includegraphics[width=\linewidth]{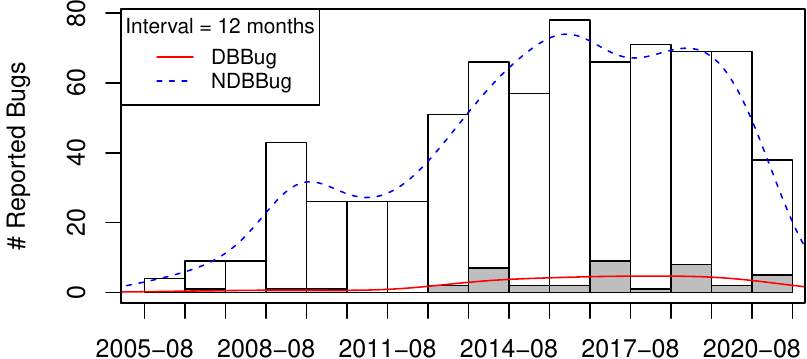}
		\caption{\texttt{Openfire}}
	\end{subfigure}
	\newline
	
	\begin{subfigure}[b]{0.32\textwidth}
		\centering
		\includegraphics[width=\textwidth]{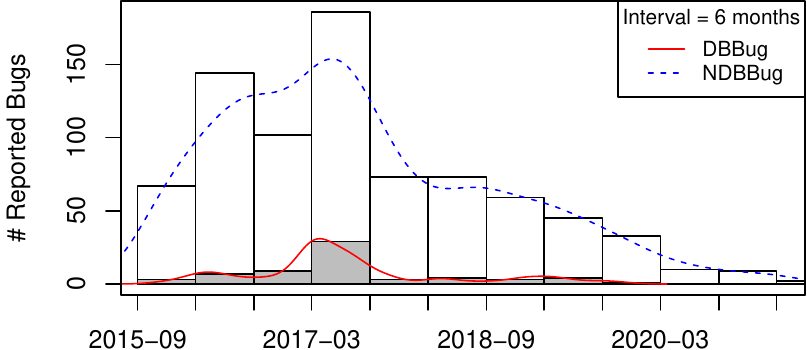}
		\caption{\texttt{ADempiere}}
	\end{subfigure}
	\begin{subfigure}[b]{0.32\linewidth}
		\includegraphics[width=\linewidth]{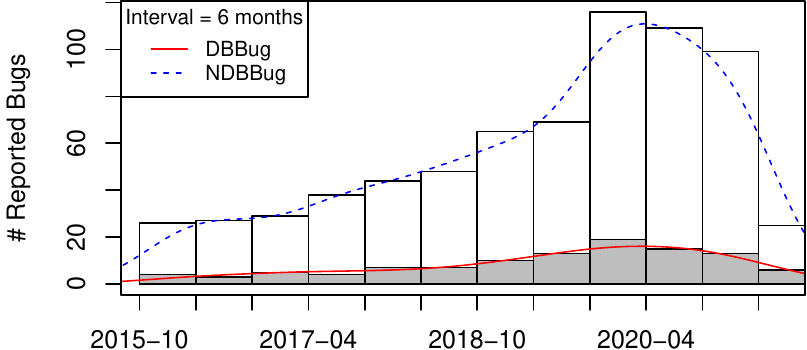}
		\caption{\texttt{DBeaver}}
	\end{subfigure}
	\begin{subfigure}[b]{0.32\linewidth}
		\includegraphics[width=\linewidth]{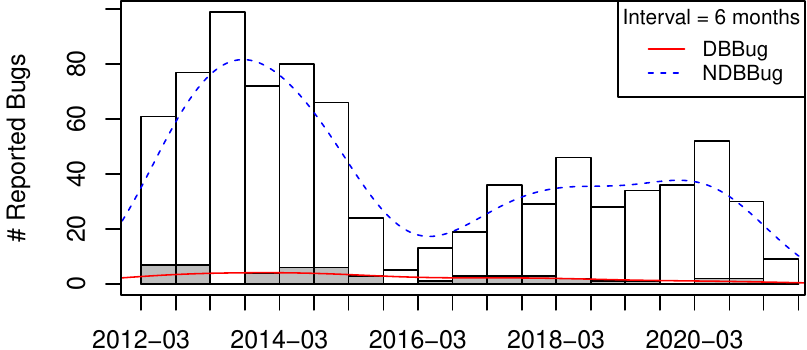}
		\caption{\texttt{dotCMS}}
	\end{subfigure}
	
	\centering
	\begin{subfigure}[b]{0.32\linewidth}
		\includegraphics[width=\linewidth]{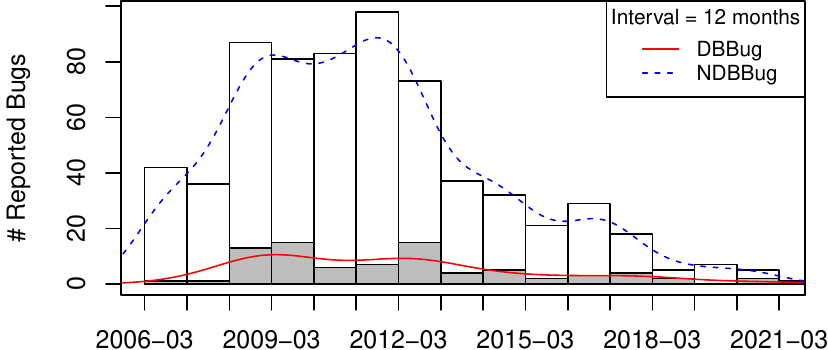}
		\caption{\texttt{OpenMRS}}
	\end{subfigure}
	
	\caption{
		The trend of reported database access bugs (DBBug) and non-database access bugs (NDBBug) across the studied applications. The reported bugs are aggregated at a fixed time interval according to their reporting time in each application.}
	\label{fig:frequency_bugs}
  \vspace{-0.2cm}
\end{figure*}

\phead{Results.} We find that the number of database and non-database access bugs share a similar trend. Figure \ref{fig:frequency_bugs} shows the trend of reported bugs during the study period. The $ x $-axis represents the time interval, and the $ y $-axis represents the number of reported bugs. For better visualization, we aggregate the reported bugs at a fixed time interval according to their reporting time in each application (e.g., interval = 3 months in \texttt{BroadleafCommerce}) and count the number of reported bugs within every time interval. 
\REM{\peter{need to describe the figures. What do the figures show? What is the axis? What are the boxes and dotted lines? }} 
We see that database access bugs are reported throughout the study period. Although sometimes more database access bugs are reported, not all of them are reported during a specific time (e.g., when the application is first released). This finding indicates that database access code maintenance~\cite{Qiu_2013_Empirical_Analysis_of_Co_Evolution_of_Schema_and_Code_in_Database_Applications, Chen_2016_An_Empirical_Study_on_the_Practice_of_Maintaining_Object_Relational_Mapping_Code_in_Java_System, Meurice_2016_Program_Inconsistencies_under_Database_Schema_Evolution} is a continuous process for database-backed applications and requires continuous attention.
We also use the density curves (i.e., the smooth lines) to fit the distribution of reported bugs. For each application, we find that both database access bugs and non-database access bugs have a similar trend - the density value increases at the beginning, reaches peaks, and decreases thereafter. 

We further apply correlation analysis (i.e., Spearman's rank correlation) to verify if there is a consistent trend between the number of reported database and non-database access bugs. We calculate the Spearman's rank correlation coefficient $ r_{s} $ between the number of reported database and non-database access bugs based on two time intervals (i.e., 3 months and 6 months) and report the results of $ r_{s} $ in Table \ref{tab:correlation_bugs_occurrence}. We find that $ r_{s} $ values of four studied applications are between 0.7 to 0.9 (\textit{strong}) while $ r_{s} $ values of two studied applications are between 0.4 to 0.6 (\textit{moderate}). The results indicate that the number of reported database access and non-database access bugs have a \textit{moderate} to \textit{strong} correlation: increasing or decreasing at the same time.

%
%

\begin{table}[ht]
	\caption{Spearman's rank correlation ($ r_{s} $) between the number of reported database and non-database access bugs across the study period. The reported bugs are aggregated using two time intervals (i.e., 3 months and 6 months).}
	\label{tab:correlation_bugs_occurrence}
	\centering
	\scalebox{0.73}{
		\begin{tabular}{lc|c}
			\toprule
			\multirow{3}{*}{\textbf{Application}} & 
			\multicolumn{2}{c}{\textbf{$ r_{s} $}}
			\\\cmidrule(l){2-3} 
			
			&\textbf{Interval = 3 months}&\textbf{Interval = 6 months}\\
			\midrule
			
			\texttt{BroadleafCommerce}  &0.9&0.9\\ 
			\texttt{metasfresh}  		&0.9&0.9\\ 
			\texttt{Openfire}			&0.5&0.7\\ 
			\texttt{ADempiere}			&0.7&0.9\\
			\texttt{DBeaver}			&0.9&0.9\\ 
			\texttt{dotCMS}             &0.4&0.5\\ 
			\texttt{OpenMRS}			&0.5&0.5\\   
			
			\bottomrule
	\end{tabular}}
  \vspace{-0.2cm}
\end{table}

\phead{Discussion.} We find that the number of reported database and non-database access bugs share a similar trend. Hence, we further investigate whether the reported database and non-database access bugs occur at similar locations by examining their modified files in bug fixing commits. Table \ref{tab:issues_unique_files} shows the number of unique modified files in bug fixing commits between database access bugs and non-database access bugs. The database access bugs occur in a small range of source code files (the number of modified files is between 33 $\sim$ 182), while the non-database access bugs occur in a large range of source code files (the number of modified files is between 452 $\sim$ 2,946). In particular, the percentage of modified files for database access bugs across modified files for all bugs is between 2\% $\sim$ 20\%. We also observe that there are 23 $\sim$ 90 common modified files (COM) shared between database access bugs and non-database access bugs. The reason is that database access code may share common files with other code (e.g., business logic code). As found in a prior study by Qiu et al.~\cite{Qiu_2013_Empirical_Analysis_of_Co_Evolution_of_Schema_and_Code_in_Database_Applications}, database schema co-evolves with the application code, which may explain the correlation between the number of reported bugs. However, our findings show that database access bugs may occur in different source code files compared to non-database access bugs. In other words, database access bugs require more research to understand their root causes and attention from the research community. In the next RQ, we manually study database access bugs to understand and create categories of the root causes.

\begin{table*}
	\caption{The number of unique modified files in bug fixing commits for reported database access bugs (DBBug) and non-database access bugs (NDBBug), the percentage (Pct.) of modified files for DBBug across modified files for all bugs (i.e., DBBug and NDBBug), and the number of common modified files (COM) shared between DBBug and NDBBug.  \REM{\peter{also, try to show the percentage of common files across all modified files, we want to show that developers change very different sets of files}}}
	\label{tab:issues_unique_files}
	\centering
	\setlength{\tabcolsep}{3pt}
	\scalebox{0.68}{
		\begin{tabular}{l|ccc|ccc|ccc|ccc|ccc|ccc|ccc}
			\toprule
			\multirow{3}{*}{\textbf{Type}} & 
			\multicolumn{3}{c}{\tabincell{l}{\textbf{\texttt{Broadleaf-}} \\ \textbf{\texttt{Commerce}}}} &
			\multicolumn{3}{c}{\texttt{\textbf{metasfresh}}} &
			\multicolumn{3}{c}{\texttt{\textbf{Openfire}}}  &
			\multicolumn{3}{c}{\texttt{\textbf{ADempiere}}} &
			\multicolumn{3}{c}{\texttt{\textbf{DBeaver}}} &
			\multicolumn{3}{c}{\texttt{\textbf{dotCMS}}} &
			\multicolumn{3}{c}{\texttt{\textbf{OpenMRS}}}
			\\\cmidrule(l){2-22}

			&\#&Pct.&\#COM
			&\#&Pct.&\#COM
			&\#&Pct.&\#COM
			&\#&Pct.&\#COM
			&\#&Pct.&\#COM
			&\#&Pct.&\#COM
			&\#&Pct.&\#COM
			\\\midrule
			
			\tabincell{l}{DBBug\\NDBBug}  
			&\tabincell{r}{119\\645}&16\%&64
			&\tabincell{r}{72\\2946}&2\%&36
			&\tabincell{r}{33\\452}&7\%&23
			&\tabincell{r}{63\\772}&8\%&31
			&\tabincell{r}{182\\735}&20\%&88
			&\tabincell{r}{133\\704}&16\%&33
			&\tabincell{r}{149\\664}&18\%&90
			\\
			
			\bottomrule
	\end{tabular}}
\vspace{-0.2cm}
\end{table*}



\rqbox{Database access bugs are reported throughout the development history of the applications. While the number of reported database and non-database access bugs share a similar trend, their modified files in bug fixing commits are different. This implies that they are not necessarily co-located and database access bugs may have different causes and fixes as compared to non-database access bugs.}



%% file: texFiles/rq2.tex
\label{sec:qualitative}
\hypertarget{rq2}{}
\subsection{RQ2: What are the Root Causes of Database Access Bugs?}
\label{rq:rq2}
\phead{Motivation.} In RQ1, we find that developers modify different files when fixing database and non-database access bugs. Thus, database access bugs may have their own unique characteristics of root cause and impact. In this RQ, we manually study database access bugs to uncover their root causes. Our goal is to create categories of the root causes that may inspire future research and help practitioners avoid common pitfalls. \textit{For researchers}, the category could guide in designing research tools to improve the quality of database-backed applications. \textit{For practitioners}, the category could serve as a checklist to define test scenarios that address specific bug types in database access.

\phead{Approach.}
For the 423 database access bugs collected in Section~\ref{sec:methodology}, we manually study their bug reports, comments, and commits to uncover their root causes. Our manual study involves three phases:

\newcounter{first}
\setcounter{first}{1} 
\newcounter{second}
\setcounter{second}{2} 
\newcounter{third}
\setcounter{third}{3}

\noindent\textit{\underline{Phase \Roman{first}.}} Two authors of the paper (A1 and A2) independently derived an initial list of the causes by manually inspecting the title, description, commit message, comment, and code change of each bug report.

\noindent\textit{\underline{Phase \Roman{second}.}} A1 and A2 unified the derived causes and compared the assigned cause for each database access bug. Any disagreement was discussed until reaching a consensus. The inter-rater agreement of the cause of bugs has a Cohen’s kappa~\cite{Cohen_1960} of 0.83, indicating an almost perfect agreement~\cite{Richard_1977_Measurement_of_Observer_Agreement}. To encourage replication of our results, we have made the dataset available online~\cite{Replication_Package}.

\noindent\textit{\underline{Phase \Roman{third}.}} We further grouped the database access bugs into five categories based on the stage when the bug occurs in the process of accessing the database: \textit{SQL Queries}, \textit{Schema}, \textit{API}, \textit{Configuration}, and \textit{SQL query result}.

\phead{Results.}
Table \ref{tab:Categories_database_related_issues} summarizes the manually-uncovered root causes of the database access bugs and the number of bug instances. Below, we discuss each category in detail. 

\begin{table*}
	\centering
	\caption{Categories of the root causes of database access bugs.}
	\label{tab:Categories_database_related_issues}
	\setlength{\tabcolsep}{2.5pt}
	\scalebox{0.65}{
	\begin{tabular}{p{1mm}p{5.2cm}p{13.8cm}r}
		\toprule
		\multicolumn{2}{l}{\textbf{Root Cause}} & \textbf{Description}  &\textbf{\# Bugs}  \\ 
		\midrule
			
		\multicolumn{2}{l}{\bf SQL Queries}  &  & \textbf{169~(40\%)}\\
		& SQL syntax error				&The SQL query contains SQL syntax error, which causes runtime exceptions. &108   \\
		& SQL logic error				&The SQL query contains incorrect business logic (e.g., incorrect condition in the \texttt{WHERE} clause), where the returned results are not what the developers expected.   &45   	\\		
		& SQL query is incompatible with some DBMSs				&The syntax of the SQL query is not compatible with some database management systems that the application promises to support. &11   	\\
		& Invalid user-defined function	&The SQL query calls a user-defined function written in PL/SQL (i.e., stored procedure) which is missing or compiled with errors.&3	  	\\
		& Error while converting data types 		& The SQL query fails when trying to convert from one data type to another. &2	  	\\
		\midrule

		\multicolumn{2}{l}{\bf Schema}  &  & \textbf{95~(22.5\%)}\\
		&Violation of database constraint &The SQL query violates database schema constraints such as not-null, foreign key, and primary key constraint.&46\\
		&Non-existent table/column			&The table/column specified in the SQL query does not exist in the database.&20		\\
		&Poor schema design 				&The design of the database scheme needs improvement or is incorrect.&11		\\ 
		&Invalid/unexpected column value 				&The SQL query inserts or updates a table column with invalid/unexpected value (e.g., the size of the value is larger than the specified column size in the DBMS).&9		\\   
		&Invalid modification of schema 	&The SQL statement modifying the table schema is invalid (e.g., duplicate values exist before creating a unique index constraint on the table columns). &9		\\
		\midrule
		
		\multicolumn{2}{l}{\bf API}  &  & \textbf{92~(21.7\%)}\\
		&Incomplete/invalid object values 		&Some fields of the entity object are not set or are set with invalid value when trying to save or update the object to the DBMS.&21	\\ 
		
		
		&Incorrect flow of calling APIs	    &The application code calling the database access API does not conform to the expected flow. For example, the code to update an entity object never calls the save method after updating the object.        &18  \\
		

		&Inconsistent entity object state	&Calling Hibernate APIs to modify entity objects whose states are inconsistent with the action (e.g., saving an entity object that is detached from Hibernate session, for which the change will not be reflected in the DBMS).	&16  \\
		&Bugs in database access APIs                     	&There are bugs in the JDBC driver or the Hibernate framework.&9 \\
		&Missing exception handling         &There is no proper exception handling when calling database access APIs (e.g., missing try-catch block).  &8  \\
		&Incorrect parameter     	&The database access API is called with missing parameters or invalid arguments.	&7  \\
		&Hibernate proxy misuse &The Hibernate proxy object, representing a lazily loading object, is accessed incorrectly (e.g., accessing non-initialized fields of a proxy object directly). 		   &6 	\\
		&Transaction misuse		   &Missing or using transaction incorrectly (e.g., missing annotation @Transactional when calling database access APIs).	&4  \\
		&Inefficient API call				&The database access API call retrieves too much unnecessary data (e.g., retrieving all fields of an object while some of them are never used in the application).	&3	\\  
		\midrule


		\multicolumn{2}{l}{\bf Configuration}  & {\REM{\peter{some of the issues don't sound like ``bugs''}\wei{Done by deleting two type of causes.}}}  & \textbf{38~(9\%)}\\  
		&Incorrect database connection    			    &The configuration of the database connection is incorrect so that the connection to the DBMS cannot be established. &16  \\
		&Incompatible database driver version   			&The database driver version is incompatible with the database version.&12  \\
		&Incorrect ORM configuration                   &Bugs in ORM configuration, such as having a typo in ORM annotations that causes unexpected database access behavior.&10  \\
		\midrule

		\multicolumn{2}{l}{\bf SQL Query Result}  &  & \textbf{29~(6.9\%)}\\ 
		&Incorrect entity object conversion  &Database-returned results are converted to entity objects incorrectly (e.g., fields mismatch between the returned database record and the entity object).&15 \\
		&Cache misuse   			&Developers use the cache of database records incorrectly (e.g., the cache is cleared unintentionally and causes performance bugs). &10  \\
		&Missing cache     				&Developers do not add the needed cache for some database tables, which causes performance bugs.&4  \\
		\bottomrule
	\end{tabular}}
  \vspace{-0.2cm}
\end{table*}

\phead{- SQL Queries (169/423, 40\%). } Database-backed applications access DBMS data by issuing SQL queries (i.e., either manually constructed by developers or automatically generated by ORM frameworks) to DBMSs. Since the compiler cannot capture errors in the SQL query during compile time, any errors in the SQL query may return unexpected results or even runtime exceptions that may cause the application to crash. 
We find that, among all the bugs related to SQL queries, \textit{syntax error in SQL queries} is the most common root cause (108/169, 63.9\%). In this case, the SQL query issued to DBMSs violates the SQL syntax rule, causing runtime exceptions. These bugs usually happen when developers make a typo in the SQL query or fail to generate the criteria as expected in the SQL query.
For example, \texttt{ADempiere} \#2494 reports a runtime exception due to the syntax error as follows:  
\begin{lstlisting}[frame=single, language=SQL,  escapechar=|, basicstyle={\scriptsize\ttfamily}]
Query.list: SELECT Gender, ... FROM C_BPartner WHERE ( |\colorbox{lightred}{AND}| C_BP_Group_ID=103) AND ... [76]
org.postgresql.util.PSQLException: ERROR: syntax error at or near "AND"
\end{lstlisting}
The SQL keyword \texttt{AND} (as highlighted in red) violates the SQL syntax. \texttt{AND} should be used between two conditions and not immediately after the keyword \texttt{WHERE}. 
The SQL query was generated based on some developer-specified criteria to find specific records in the database table. The problematic SQL query was caused by some untested criteria, which resulted in generating incorrect logical operators (i.e., \texttt{AND}, \texttt{OR}, and \texttt{NOT}) to combine conditions in the \texttt{WHERE} clause.

The second most common root cause is \textit{logic error in SQL queries} (45/169, 26.6\%). Although the SQL query issued to DBMSs may be syntactically correct, the returned result may not be what the developers expected. These bugs usually happen when developers partially understand the business requirements of the underlying data query. For example, there may be a logical error (e.g., missing conditions) in the \texttt{WHERE} clause of the SQL query (e.g., \texttt{BroadleafCommerce} \#586), which leads to unexpected query results. We also find cases where the \textit{SQL queries are not compatible with some DBMSs} that the application promises to support (11/169, 6.5\%). In some cases (3/169, 1.8\%), we find that the \textit{SQL query calls invalid user-defined functions}, which is missing (e.g., non-existent) 
or compiled with errors (e.g., \texttt{ADempiere} \#828). Hence, the SQL query causes runtime exceptions, leading to no query results. Finally, in 2/169 (1.2\%) cases, we find that there are \textit{data conversion errors in SQL queries}. For example, in \texttt{Adempiere} \#1174, the SQL query calls the \texttt{COALESCE()} function which fails to implicitly convert the data type from \texttt{VARCHAR} to \texttt{NVARCHAR2}. Developers fixed these bugs by using the target data type directly to avoid erroneous implicit conversion. 

Overall, based on our manual observation, many SQL-related bugs may be revealed if developers have proper test cases in place. For example, if test cases cover all the SQL queries, then the test cases should be able to capture SQL syntax errors. Moreover, some logic errors in the SQL queries (e.g., the returned result is unexpected) may also be revealed if there are comprehensive test cases. After manually checking the test cases, we find that many problematic SQL queries are not fully covered by test cases. One possible way to improve the quality of database access code is by introducing SQL query coverage as code coverage criteria to measure how well the database access code is tested.
For researchers, our findings call for automatic test generation tools~\cite{Pan2014_Guided_Test_Generation_for_Database_Applications,Emmi_2007_Dynamic_Test_Input_Generation_for_Database_Applications, Castelein_2018_Search_Based_Test_Data_Generation_for_SQL_Queries, Arcuri_2020_Handling_SQL_Databases_in_Automated_System_Test_Generation} that can cover SQL queries in database access. 


\rqbox{Any errors in SQL queries issued to DBMSs may cause unexpected query results or even runtime exceptions. Our manual analysis finds that these problematic SQL queries are not fully covered by test cases. Future studies should emphasize the development of test generators that target SQL query coverage.}


\phead{- Schema (95/423, 22.5\%). } The database schema defines how data is structured in the database (e.g., tables and data types) and how data records inside the database relate to each other (e.g., \texttt{foreign key}). When database-backed applications access the data, the DBMS receives the SQL query and executes it by querying the data defined by the database schema. Hence, any issues related to database schema may return unexpected results or cause runtime exceptions. We find that, among all the bugs related to the database schema, \textit{violation of database constraint} is the most common root cause (46/95, 48.4\%). In these cases, the SQL query violates one or more of the common database constraints: \texttt{foreign key}, \texttt{unique}, \texttt{not-null}, and \texttt{primary key} constraints, which causes an exception to occur. The violation of database constraints usually happens if developers do not handle corner cases or exceptions properly when they try to persist the data in database tables.
As the database constraints are configured in the DBMS, the developers may not know their existence or fail to validate if the data complies with the database constraints in the application code. For example, \texttt{Openfire} \#692 reports a runtime exception due to the violation of \texttt{not-null} database constraint when an SQL query tries to insert the value of \texttt{NULL} into a column that does not allow null value. The code snippet is shown as follows:

\begin{figure}[htb]
\centering
\begin{adjustbox}{scale=0.85}
\begin{subfigure}[b]{0.75\textwidth}
\begin{lstlisting}[frame=single, language=SQL, numbers=left, escapechar=|, breaklines=true, basicstyle={\scriptsize\ttfamily}]
public class XMPPServer {
    ...
    try {
        host = InetAddress.getLocalHost().getHostName();
    } catch (UnknownHostException ex) {
        Log.warn("Unable to determine local hostname.", ex);
        host = "127.0.0.1";
    }
}

public class DefaultSecurityAuditProvider {
    ...
    String LOG_ENTRY = "INSERT INTO ofSecurityAuditLog (msgID,username,entryStamp,summary,node,details) VALUES(?,?,?,?,?,?)";
    PreparedStatement pstmt = con.prepareStatement(LOG_ENTRY);
    ...
    pstmt.setString(5, XMPPServer.getInstance().getServerInfo().getHostname());
    pstmt.executeUpdate(); //execute the SQL
}
\end{lstlisting}
\end{subfigure}
\end{adjustbox}
 \vspace{-0.4cm}
\end{figure} 

The value of \texttt{host} (i.e., host name) is initialized in class \texttt{XMPPServer} (Line 4). In case \texttt{UnknownHost\\Exception} happens, developers try to assign an IP address (i.e., \texttt{127.0.0.1}) to the host name in the \texttt{catch} block (Line 4). Then, in class \texttt{DefaultSecurityAuditProvider}, developers construct the SQL query (Line 13) and set the parameter \texttt{node} with the value of host (Line 16). However, if an uncaught exception (any exception that is not \texttt{UnknownHostException)} occurs in class \texttt{XMPPServer}, the host name would be null and executing the SQL query in Line 17 would violate the \texttt{not-null} database constraint (the corresponding column of the table for \texttt{node} should not be NULL).

The second most common root cause is \textit{non-existent table/column } in the database (20/95, 21.1\%). Database-backed applications and the underlying database evolve during the software development, which may lead to inconsistency between the SQL query and database schema. For instance, the table/column specified in the SQL query may be deleted, renamed, or not yet created (e.g., \texttt{dotCMS} \#4806) in the database. 
Another root cause is \textit{poor schema design} (11/95, 11.6\%), where the design of the database schema needs improvement or is incorrect, which may cause unexpected results. For example, missing a unique constraint on specific table columns may result in unintended duplicate table records on those columns (e.g., \texttt{dotCMS} \#5755). We also find cases where the SQL query inserts or updates a table column with invalid/unexpected value (9/95, 9.5\%). For example, the size of the value in the SQL query may be larger than the specified column size in the DBMS  (e.g., \texttt{OpenMRS} \#601). Finally, in 9/95 (9.5\%) cases, we find that the SQL statement modifying the table schema is invalid because it violates how data is structured in the database or the constraint rule. For example, in \texttt{Adempiere} \#1174 the \texttt{CREATE UNIQUE INDEX} statement fails to create the \texttt{unique} constraint as duplicate records already exist on the table columns.

We find that database access bugs caused by violation of database constraints usually happen if developers do not handle corner cases properly (i.e., untested conditions), which leads to an invalid value of persistent data that violates database schema constraints. For example, \texttt{NULL} value is invalid to be inserted into the database table if the corresponding column is configured with a \textit{not-null} database constraint. The invalid value of persistent data may be detected earlier by test cases or avoided if developers have proper validation of the data before persisting it to the DBMS.
We also find that the co-evolution of underlying database schema and code~\cite{Qiu_2013_Empirical_Analysis_of_Co_Evolution_of_Schema_and_Code_in_Database_Applications, Wang_2019_Synthesizing_Database_Programs_for_Schema_Refactoring} in database-backed applications may lead to violation of database schema in SQL queries (e.g., \textit{non-existent table/column}). For example, the table name may still remain the same in SQL queries when it has been modified in the DBMS. Our findings suggest that developers should consider the underlying database schema when generating SQL queries and keep track of the database schema evolution to update the SQL query. Our findings also call for an automatic mechanism to maintain the consistency between SQL queries and database schema when the schema evolves.


\rqbox{SQL queries with invalid values of persistent data may violate database schema constraints and SQL queries may also violate database schema due to the database schema evolution, causing runtime exceptions. Future studies may provide support for the development and maintenance on SQL queries regarding database schema.}


\phead{- API (92/423, 21.7\%). } In database-backed applications, developers call database access APIs (e.g., ORM APIs) to access database data. Bugs are introduced if developers partially understand the assumptions made by the rich set of APIs or use the API in a way that does not conform to the business logic of applications. While most issues are related to incorrect or inefficient usages of APIs (90.2\%), there are still some issues related to the API itself (i.e., \textit{bugs in database access APIs}, 9/92, 9.8\%). We find that, among all the bugs related to database access API usage, \textit{incomplete/invalid values in the entity object} is the most common root cause (21/92, 22.8\%).  
These bugs usually happen when developers call database access APIs to persist the entity object with many fields, of which developers forget to set values of some fields (e.g., \texttt{OpenMRS} \#3337) or set some fields with invalid values, resulting in incorrect records in database tables.
The second most common root cause is \textit{business logic error} when calling database access APIs (18/92, 19.6\%). In this scenario, the application code calling the database access API does not conform to the application's business logic which may cause unexpected records in database tables. For example in \texttt{BroadleafCommerce} \#1538, the application intends to set the \texttt{payment} status in the database to \textit{archived}. However, developers never explicitly call the database API to update the corresponding database record, which causes incorrect \texttt{payment} status. The third most common root cause is \textit{inconsistent entity object state} when calling database access APIs (16/92, 17.4\%). Developers may misuse the Hibernate session to modify the entity object, causing Hibernate exception \texttt{NonUniqueObjectException}, which means that the developer tries to associate two different entity objects with the same identifier value (i.e., primary key), in the scope of a single session. For example, the developer may try to save an entity object that is detached from Hibernate session, when a different object with the same identifier value is already associated with the session (e.g., \texttt{OpenMRS} \#3728).

The next two most common root causes are \textit{bugs in database access APIs} (9/92, 9.8\%) and \textit{missing exception handling} when calling database accessing APIs (8/92, 8.7\%). 
For example, \texttt{DBeaver} \#6554 reports unexpected query results caused by a bug in the JDBC driver for SQL Server (\texttt{mssql-jdbc} \#969~\cite{mssql-jdbc-issue}). We also find cases (7/92, 7.6\%) where the database access API is called with \textit{incorrect parameter}, missing parameters or invalid arguments (e.g., \texttt{null} argument in \texttt{BroadleafCommerce} \#153). In some cases (6/92, 6.5\%), we find that developers \textit{misuse Hibernate proxy} where the Hibernate proxy, representing lazily loading object, is accessed incorrectly. The non-initialized fields of the proxy object are accessed directly by developers (e.g., \texttt{OpenMRS} \#3340) instead of calling the associated getter method which enforces Hibernate to initialize fields by querying the DBMS. In some cases (4/92, 4.3\%), we find that developers \textit{misuse the transaction}, which may cause data integrity bugs in the database (e.g., missing transaction in \texttt{BroadleafCommerce} \#330). Finally, in (3/92, 3.3\%) cases, we find that developers call the database API to retrieve too much unnecessary data, which may cause performance bugs (e.g., \texttt{BroadleafCommerce} \#762).

We find that, when calling database APIs, many database access bugs are related to the discrepancy between database records and objects in object-oriented languages. Although ORM frameworks provide developers with a conceptual abstraction for mapping the database records to objects, any errors in entity objects may cause incorrect records in database tables (e.g., \textit{incomplete/invalid values in the entity object}). 
We also find that developers may have difficulties managing the entity object when calling database access APIs (e.g., \textit{inconsistent entity object state}) due to the complexity and impedance mismatches~\cite{Chen_2016_Finding_and_Evaluating_the_Performance_Impact_of_Redundant_Data_Access} of the object-relational mapping. Simplifying the complexity of database access APIs may help reduce database access bugs. Future studies may also provide support to developers in using database access APIs. For example, an entity object checker may help developers detect incomplete values in the entity object when calling database access APIs.


\rqbox{When calling database APIs, many of the issues are related to incorrect API usage or API anti-patterns. Future studies may provide support for developers in using database access APIs.}

\phead{- Configuration (38/423, 9\%). }
In database-backed applications, developers need to set up various configurations (e.g., database connection) before accessing the database. Incorrect configurations may cause errors or unexpected behaviors when accessing the database. We find that, among all the database access bugs related to configuration, \textit{incorrect database connection} is the most common root cause (16/38, 42.1\%). The database connection is configured as a URL that contains information such as where to search for the database (i.e., the host name and port number of the node hosting the DBMS) and the name of the database to connect to. The information in the database connection URL may be incorrectly configured (e.g., \texttt{DBeaver} \#9382), causing the connection error to the DBMS. The second most common root cause is \textit{incompatible database driver version} (12/38, 31.6\%), which may lead to database access failure. For example, some data types for a specific DBMS release may be changed and not supported by the database driver (e.g., \texttt{Openfire} \#759), causing runtime exceptions. We also find cases where the \textit{ORM configuration is incorrect} (10/38, 26.3\%). Since the ORM frameworks automatically convert entity objects to/from the corresponding database record based on the configuration, incorrect configuration of ORM may lead to unexpected ORM framework behaviors. For example, in \texttt{BroadleafCommerce} \#497, a duplicate annotation \texttt{JoinTable} is configured on the entity class \texttt{OfferCode}, which incorrectly adds additional duplicate records in the database table.

We find that developers may incorrectly set up configurations before accessing the database, which may cause errors or unexpected behaviors when accessing the database. We also find that developers may forget to update the configurations (e.g., database driver version) when the code or DBMS evolves. Our findings indicate that managing the configuration is a continuous process during the development and evolution of database-backed applications. Developers may benefit from tools that can detect incorrect configurations automatically. For example, an ORM configuration checker may help developers detect duplication annotation in ORM configurations.


\rqbox{Incorrect configurations may cause errors or unexpected behaviors when accessing the database. Developers should develop automated tests that continuously verify various configurations. Future studies may also work on tools to help developers detect incorrect configurations automatically.}

\phead{- SQL Query Result (29/423, 6.9\%). }
Database-backed applications often convert the data records returned by the DBMS into objects in object-oriented programming languages. For frequently-queried data in the DBMS, they also store the corresponding objects in the cache. We find that, among all the bugs related to SQL query results, \textit{incorrect entity object conversion} is the most common root cause (15/29, 51.7\%). These bugs are caused by incorrect conversion from database-returned results into entity objects and usually happen when developers mismatch the fields between them (e.g., \texttt{Openfire} \#664), thereby causing inconsistency between entity objects in applications and database records. We also find cases where developers \textit{misuse the cache} for database table records (10/29, 34.5\%). For example in \texttt{dotCMS} \#5553, developers incorrectly clear the entire cache instead of the corresponding cache needed after updating the data and saving it in the DBMS, which may cause performance bugs due to the unnecessary cache updates. Finally, in 4/29 (13.8\%) cases, we find that developers did not use cache to store frequently-queried data in database tables (e.g., \texttt{dotCMS} \#6288), which causes significant data retrieval overhead.

We find that database access bugs caused by \textit{incorrect entity object conversion} usually occur when calling JDBC APIs to access the database. When using JDBC APIs, developers have to extract the SQL query results and convert the values to corresponding fields of the entity object. In contrast, when using ORM frameworks, the conversion is done automatically after developers configure the mapping between entity classes and database tables. While caching is a common way to improve the performance of database access, we also find that misuse of the cache may introduce performance bugs. Future studies may propose tools to help developers use caching frameworks when accessing the database~\cite{Chen_2016_CacheOptimizer}.

\rqbox{When retrieving SQL query results, developers may incorrectly convert the database records returned by the DBMS into entity objects in applications, causing inconsistency between them. Developers may also misuse the cache which causes performance bugs, calling for tools to help developers use caching frameworks more intelligently when accessing the database. }

%% file: texFiles/rq3.tex
\hypertarget{rq3}{}
\subsection{RQ3: How do Categories of Database Access Bugs Prevail with Different Database Access Technologies?}
\label{rq:rq3}

\phead{Motivation.}
In RQ2, we analyzed the root causes of database access bugs and grouped them according to several categories. In general, database accesses leverage two widely used technologies: (i) SQL queries (e.g., JDBC), and (ii) ORM (e.g., Hibernate). These database access technologies have unique design principles and goals. For example, rather than constructing SQL queries in the database access code, ORM enables developers to manipulate persistent data as if it is in-memory objects~\cite{Yan_2017_Understanding_Database_Performance_Inefficiencies_in_Real_World_Web_Applications}. Thus, studying the bugs that are related to the two technologies allows us to further understand their maintenance challenges. In this RQ, we investigate how different categories of database access bugs prevail with these two technologies. 
Our findings may guide future studies to provide support for developers in maintaining database-backed applications that use different technologies.

\phead{Approach.}
We further analyze the database access bugs that we studied in RQ2 and classify the bug reports by the technology (i.e., JDBC or Hibernate) used to access the DBMS. We perform a manual inspection of the source code associated with the bug fix to verify the usage of the technologies. For each bug category, we compare the percentage of database access bugs that is related to each different technology.

\REM{
\begin{figure}
	\centering
	\includegraphics[width=0.8\linewidth]{./image/distribution_types}
	\caption{Distribution of database access bug types across applications. ``Broadleaf.'' is abbreviation of ``BroadleafCommerce''. \shouvick{The SQL queries bar color is not looking good. Please change!}}
	\label{fig_distribution_types}
\end{figure}
}

\REM{
\phead{Results.} We find that SQL Queries bugs occur most frequently in three applications \texttt{Openfire}, \texttt{ADempiere} and \texttt{DBeaver}. Schema bugs occur most frequently in two applications \texttt{metasfresh} and \texttt{dotCMS} while API bugs occur most frequently in two applications \texttt{BroadleafCommerce} and \texttt{OpenMRS}. In total, three bug types (i.e., SQL Queries, Schema and API) cover more than 85\% of bugs across all studied applications. }

\begin{figure}[h]
	\centering
	\includegraphics[width=0.5\linewidth]{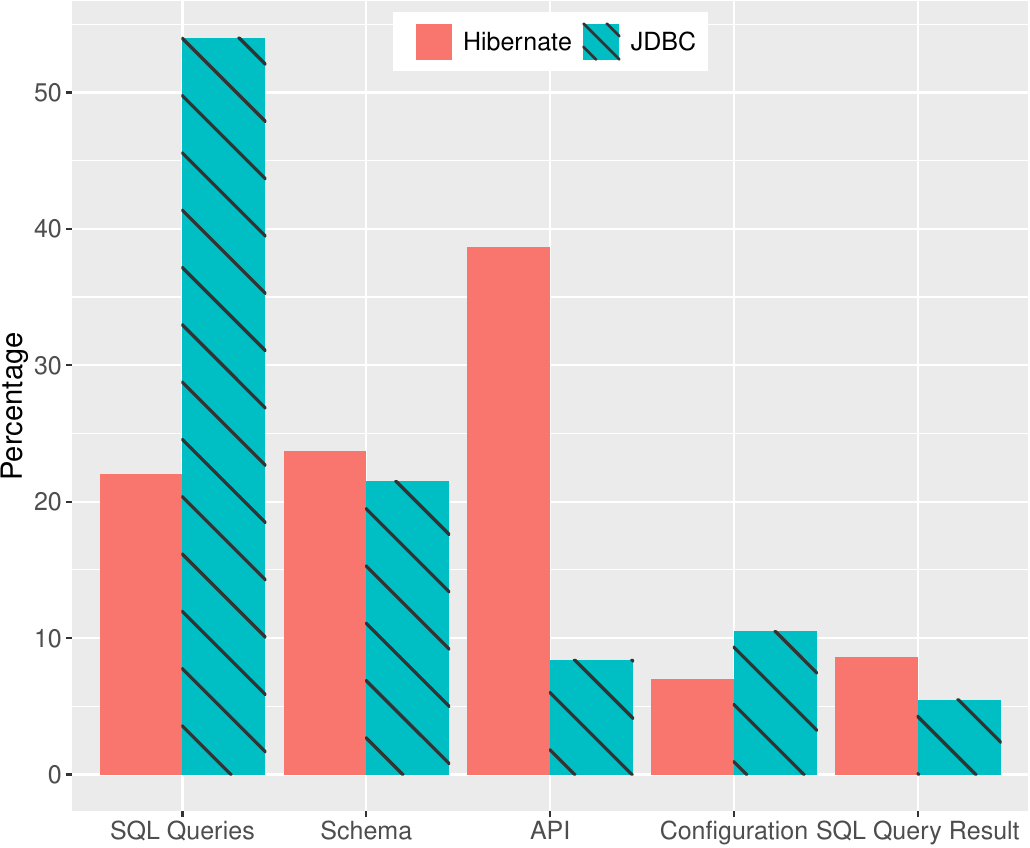}
	\caption{Distribution of the categories of database access bugs that occur in JDBC and Hibernate database-backed applications.}
	\label{fig_distribution_types_freamwork}
     \vspace{-0.2cm}
\end{figure}

\phead{Results.}
Among all the studied applications, \texttt{metasfresh}, \texttt{Openfire}, \texttt{ADempiere}, and \texttt{DBeaver} use JDBC while \texttt{BroadleafCommerce}, \texttt{dotCMS}, and \texttt{OpenMRS} use Hibernate to access the database. Accordingly, 237 database access bugs are related to JDBC, and 186 are related to Hibernate. Figure \ref{fig_distribution_types_freamwork} compares the distribution of database access bug categories between JDBC and Hibernate. We find that \textit{SQL queries}, \textit{Schema}, and \textit{API} bugs cover (356/423, 84.2\%) database access bugs.



\phead{- SQL queries bug (128/237, 54\%) is the most frequent issue when using JDBC, while API bug (72/186, 38.7\%) is the most frequent issue when using Hibernate.} The possible reason is that developers manually construct the SQL queries when using JDBC, while Hibernate generates the SQL queries automatically which makes it less prone to errors in SQL queries (e.g., SQL syntax errors). In terms of API bugs, the advanced Hibernate features (e.g., Hibernate session and Hibernate proxy) make Hibernate APIs more complex to use compared to JDBC APIs. We find that, when using Hibernate, most of the API bugs happen when developers call database access APIs to persist the entity objects (e.g., updating the value of the entity object to the corresponding database record). Our findings suggest that developers should pay more attention to constructing the SQL queries when using JDBC and pay more attention to API usage, especially persistent APIs, when using Hibernate. 

\phead{- There are many SQL query bugs (128/237, 54\%) when using JDBC, while there are still SQL query bugs (41/186, 22\%) when using Hibernate.} As discussed in RQ2, most of the SQL query bugs are caused by \textit{syntax error in SQL queries}. Although Hibernate automatically generates the SQL queries executed by DBMSs, it provides APIs to use Hibernate Query Language (HQL) queries~\cite{HQLandJPQL}. Hibernate may also generate problematic SQL queries if there are errors in HQL queries manually constructed by developers (e.g., \texttt{OpenMRS} \#5359). Hence, using Hibernate may still result in SQL query bugs. 

\phead{- There are schema bugs when using JDBC (51/237, 21.5\%) and Hibernate (44/186, 23.7\%).} 1) When using JDBC, schema bugs are mainly caused by \textit{Non-existent table/column} (18/51, 35.3\%), mostly due to the co-evolution of underlying database schema and code (as discussed in RQ2). In contrast, schema bugs caused by \textit{Non-existent table/column} only account for (2/44, 4.5\%) when using Hibernate. The possible reason is that, when database schema evolves (e.g., the table column name changes), developers only need to modify the mapping between entity/table once when using Hibernate, but have to manually modify all corresponding SQL queries when using JDBC. Developers using JDBC may benefit a lot from automatic tools to help maintain the consistency between SQL queries and database schema. 2) When using Hibernate, schema bugs are mainly caused by \textit{violation of database constraint} (31/44, 70.5\%). Hibernate provides built-in constraints~\cite{Built_in_constraints} on entity fields to help developers prevent \textit{violation of database constraint}. For example, the annotation \textsf{@NotNull} in Hibernate declares a field to be not-null. If the field value is null, Hibernate will not execute any SQL statements and prevents storing null values in the underlying database, which avoids the violation of database constraint \texttt{not-null}. However, Hibernate does not provide corresponding built-in constraints for other database constraints such as \texttt{foreign key} (e.g., \texttt{BroadleafCommerce} \#678), \texttt{primary key}, or \texttt{unique constraint}. Developers of Hibernate framework may provide more built-in constraints in the future to help developers deal with common database constraints~\cite{Yang_2020_Yang_Managing_Data_Constraints}.

\phead{- There are cache issues in SQL query result bugs when using JDBC (5/13, 38.5\%) and Hibernate (9/16, 56.3\%)}
We find that, when using JDBC and Hibernate, some SQL query result bugs are related to the cache, which may cause performance issues. Future studies of performance in database-backed applications may address SQL query results when accessing the database.



		
		
		 
		
		
		

\rqbox{SQL queries, Schema, and API bugs cover 84.2\% of database access bugs across all studied applications. SQL queries bug (54\%) and API bug (38.7\%) are the most frequent issues when using JDBC and Hibernate, respectively. Hibernate cannot abstract database access completely, so many issues such as SQL query bugs and Schema bugs still exist when using Hibernate. Future studies should provide support for developers in using different database access technologies.}

%% file: texFiles/discussion.tex

\section{Discussion}
\label{sec:discussion}
We now state the actionable implications of our findings and highlight opportunities for future work.

\phead{There is a need for better support for the development and maintenance of database access code in database-backed applications.}
In RQ2 and RQ3, we find that bugs related to SQL queries or the database schema (e.g., syntax error or inconsistency with the database schema) are the most frequent category of database access bugs when using JDBC. For example, developers may make a typo in the SQL query or fail to construct the search criteria as expected in the SQL query, leading to problematic SQL queries which are not checked at compile time. 
We also find that this type of bug still exists when using Hibernate, since developers may need to use JPQL for more complex database access. 
Therefore, to assist developers with improving the quality of database-backed applications, there is a need for better tooling support to verify the SQL queries and database schema. 
For example, future research may work on tools that statically verify the syntax of SQL queries and the consistency between database schema and the SQL queries in the application code. 
Although there are some static analysis tools, such as \texttt{dbcritic}~\cite{dbcritic} and \texttt{holistic}~\cite{holistic}, that try to help detect SQL schema issues using static analysis,  these tools can only analyze specific database access frameworks (mostly only JDBC) or DBMS due to the limitation of static analysis.  
Building more generic static analysis tools that can detect errors in SQL queries will alleviate more SQL query bugs during system development. 
Future research may also work on approaches to automatically maintain database access code. Developers may benefit from approaches that automatically suggest updates to the database schema or database access code when developers modify the code -- change impact analysis. 
For example, when a developer modifies the database schema, the approach can automatically identify all the database access code (e.g., either SQL queries or ORM database access APIs) that is impacted by the change and require an update. 


\phead{Future studies should help developers better leverage ORMs such as Hibernate.} 
In RQ2 and RQ3, we find that API bug is the most frequent issue when using Hibernate. Compared to JDBC APIs, the advanced Hibernate features (e.g., Hibernate session and Hibernate proxy) make Hibernate APIs more complex to use. When calling Hibernate APIs, developers may persist entity objects with errors (e.g., \textit{incomplete/invalid object values}), which causes incorrect records in database tables. 
While there are many prior studies~\cite{Chen2014_Detecting_Performance_Anti_Patterns_for_ORM, Yan_2017_Understanding_Database_Performance_Inefficiencies_in_Real_World_Web_Applications, Yang_2018_Structure_Database_Backed_Web_Applications, Shao_2020_Database_Access_Performance_Antipatterns, 2019_Quantifying_the_Performance_Impact_of_SQL_Antipatterns_on_Mobile_Applications} focus on helping developers detect performance issues when using ORM APIs, there is limited study on the functional aspect. 
Future studies should provide support for developers in using Hibernate APIs. For example, future research may propose an entity object checker that helps developers detect incomplete values in the entity object before calling Hibernate APIs. Another possible idea is to check whether every field of the persistent entity has been initialized or set with a value. This helps ensure the validity of the database-managed objects and avoid runtime errors. 
We also find that there are many issues related to the incorrect flow of calling APIs or API parameters. Future studies can also propose approaches to help detect issues in ORM API usage.


\phead{Designing and Generating adequate test cases for database access code.} In RQ2, we find that most database access bugs cause severe problems like unexpected query results or runtime exceptions that may cause the application to crash. Our manual study finds that these bugs are usually not fully covered by test cases. For example, SQL queries with invalid values (e.g., \texttt{NULL}) of persistent data may violate database schema constraints (e.g., \textit{not-null}). These bugs occur if invalid values are generated from untested conditions. In order to improve the quality of database-backed applications, developers may consider coverage of both SQL queries (syntax and semantics) and database access API usage to measure how well the database access code is tested. Developers may also consider different types of coverage such as database coverage (e.g., coverage based on database schema, constraints, or database access code) since databases are the key components of database-backed applications. Future studies may consider this coverage as code coverage criteria in automatic test generation tools for database-backed applications. For example, future studies may use database coverage to guide test case generation using search-based or fuzz-based approaches. Future studies may also propose metrics or approaches to evaluate the effectiveness of the existing test cases. For instance, there may be a need to design specialized mutation operators for database-related tests and help developers with the quality assurance of database-backed applications. 

\phead{Complementary to developers in selecting database access technologies.}
Developers often rely on two main technologies (i.e., SQL queries and ORM frameworks) to access the underlying database and discuss a lot on how to select them. We select the top 15 questions that compare Hibernate to JDBC in Stack Overflow and manually examine the answers. We find that developers often make trade-offs in the selection, mainly focusing on the strengths and limitations between JDBC and Hibernate. For example, one strength of using Hibernate is that developers do not need to write SQL queries (\textit {``Such ORMs provide the maximum level of abstraction to the point you almost never have to write SQL queries.''}~\cite{JPA_or_JDBC}). On the other hand, no explicit SQL in the source code when using Hibernate sometimes makes debugging and performance tuning difficult (\textit{``The time savings gained are easily blown away when you have to debug abnormalities resulting from the use of the ORM.''}~\cite{Java_ORM}). We also contacted the main contributors of the studied applications by inquiring why and how they selected JDBC or Hibernate in their applications. They mention that Hibernate makes their code easy to understand and modify since the application is open source and has contributors at every level from around the world. However, they also address that developers may not truly know the automatically generated SQL queries by Hibernate, which may cause major slowdowns or failures after deployment. Our findings provide complementary to developers in selecting database access technologies. For example, considering developers’ capabilities in technologies, if they have a good understanding of the entity object in Java code, they may make fewer bugs (i.e., database access bugs related to \textit{incomplete/invalid values in the entity object}  or \textit{inconsistent entity object state}) when calling Hibernate APIs.

%% file: texFiles/threats.tex
\section{Threats to Validity}
\label{sec:threats}

\phead{External Validity.}
One possible threat to external validity is the generalization of the dataset (i.e., database access bugs) we collected. To ensure that the applications we study are large enough and well maintained, we apply three criteria to select the studied Java applications based on the number of stars, the use of issue tracking systems, whether containing database-related bugs, and active maintenance activities (having more than 1,000 commits and at least 100 fixed bug reports). Based on these criteria, we ended up with seven applications, and these applications happened to use Hibernate or JDBC to access the DBMS. These applications contain thousands of lines of code (117K $\sim$ 1,610K), thousands of commits (10K $\sim$ 52K), and a pronounced development period (5 $\sim$ 16 years), across different domains such as e-commerce, ERP, and database tools. 
We do notice that metasfresh is a fork of ADempire, but the fork was done in 2015 due to the development gap compared to the latest ADempiere codebase\footnote{https://en.wikipedia.org/wiki/Metasfresh}. Since then, both applications have grown apart and there has been active development on metasfresh. As shown in Table \ref{tab:number_issues}, the studied periods of both applications are from after 2015, which cover the multiple years of development activities and database access bugs after the fork. The LOC also differs significantly between metafresh and ADempiere (879K vs. 1,610K). Therefore, due to the difference in development activities, we keep these two applications in our study.
Developers can also use these database access frameworks to interact with a wide range of underlying relational DBMSs (e.g., MySQL, PostgreSQL, or Oracle). There may be some framework-specific issues when using other database access frameworks, but many of the issues that we found in this paper are not specific to one framework (e.g., there are many bugs related to SQL query syntax or database schema). Therefore, we believe that our findings still provide important implications for both researchers and practitioners.
We acknowledge that non-relational databases such as NoSQL are becoming more popular. However, in this paper, we focus on studying the systems that use relational database management systems (e.g., MySQL or PostgreSQL). This focus is due to their established history and the relative scarcity of studies examining the characteristics of database access bugs in these environments. Future studies should consider NoSQL as it is also a critical topic, especially given the variety of NoSQL database vendors and data types (e.g., graphs or documents). 


\phead{Internal Validity.}
The main threat to the internal validity of our results could be the bias when deriving bug causes. To mitigate this threat, two authors of the paper (A1 and A2) independently inspected the bug report and commit to each bug to identify the cause. The inter-rater agreement of the cause of bugs was measured using Cohen's Kappa coefficient and the disagreements were discussed until reaching a consensus.
We considered only the database access technologies (i.e, JDBC and Hibernate) as the factors for the categories of database access bugs. However, other factors (e.g., design, coding style, and framework) may also affect the design of database access code. To mitigate this threat, we carefully examined the documentation and source code of the studied applications and found that they mainly use JDBC or Hibernate directly to access the database. 
Therefore, we believe that other factors should not significantly affect the database access and thus the occurrence of categories of database access bugs.
We only conducted our study in Java applications, but there are other database-backed applications implemented in other programming languages such as Python and Ruby. These programming languages also have various ORM frameworks available (e.g., Django, and Ruby on Rails) that may have their own unique challenges. 
Based on the literature~\cite{Chen2014_Detecting_Performance_Anti_Patterns_for_ORM, Chen_2016_Finding_and_Evaluating_the_Performance_Impact_of_Redundant_Data_Access, Chen_2016_CacheOptimizer, Yang_2018_Structure_Database_Backed_Web_Applications, Yan_2017_Understanding_Database_Performance_Inefficiencies_in_Real_World_Web_Applications}, there are many overlaps in the database access performance problems between Ruby and Java. Common issues in both Ruby and Java include retrieving more data than needed, not using batching for database access, caching, and inefficient ORM API usage. Therefore, we expect that there are some similar issues in the database-backed applications that are implemented in other programming languages, but there should also be language- and framework-specific issues. Future studies are needed to further study the issues in applications that are implemented in other programming languages and identify commonalities and differences between them.

\phead{Construct Validity.}
The construct validity of our study rests on the methodology of collecting database access bugs from the fixed bug reports of studied applications. First, for applications with fixed bug reports of more than 1,000, we conduct the study on a statistically significant sample from all fixed bug reports randomly under a 95\% confidence level and a 3\% margin of error, which may introduce tiny noise. Second, we filtered the studied bug reports by searching for database-related keywords in each bug report. The keywords were derived based on manual analysis and hence, may not be comprehensive. We may have omitted seemingly trivial database-related bugs with very little reported information.
The similarity in the trends between the database and non-database bugs may be related to the application's development process. Hence, we manually examine the code repositories, release notes, and development history to uncover the possible development process that the applications follow. Overall, we find that all the studied applications have adopted continuous integration and agile development, at least since the past decade. Most of the applications, such as DBeaver and dotCMS, have a consistent release cycle of three months. We further examine the spikes in the number of reported issues as shown in Figure~\ref{fig:frequency_bugs}, and we find that most spikes are related to having more code changes (e.g., a major release). In short, we do not find a clear connection between the software development process and the trends across the studied applications. On the other hand, we find specific reasons for some database access bugs (as discussed in RQ2). For example, \texttt{ADempiere} \#2494 reports a runtime exception due to the syntax error in the SQL query. The SQL query was generated based on some developer-specified criteria to find specific records in the database table. The problematic SQL query was caused by some untested criteria, which resulted in generating incorrect logical operators (i.e., AND, OR, and NOT) to combine conditions in the WHERE clause. Another example is \texttt{Openfire} \#692 which reports a runtime exception due to the violation of not-null database constraint when an SQL query tries to insert the value of NULL into a column that does not allow null value. The bug happens when developers do not handle corner cases properly (i.e., untested conditions). In short, these bugs may happen at any stage of the development when developers modify existing database access code, fix other bugs, or add new features. However, our finding shows that database access code is the core of database-backed applications. Although these applications may already have a comprehensive design of the database, the database access code co-evolves with the other source code and they also require continuous attention for maintenance and quality assurance. 

%% file: texFiles/related.tex
\section{Related Work}
\label{sec:related}

In this section, we discuss the related work of our paper.  


\phead{Database access issues in database-backed applications.} 
Prior works study the database access issues in database-backed applications from different perspectives, e.g., syntactic or semantic errors in SQL queries ~\cite{Brass_2004_Semantic_errors_in_SQL_queries, Ahadi_2016_Semantic_Mistakes_in_Writing_Seven_Different_Types_of_SQL_Queries}, SQL anti-patterns~\cite{Karwin_2010_SQL_Antipatterns_Avoiding_the_Pitfalls_of_Database_Programming, Arzamasova_2018_Cleaning_Antipatterns_in_an_SQL_Query_Log, Dintyala_2020_SQLCheck_Automated_Detection_and_Diagnosis_SQL_Anti_Patterns, Alshemaimri_2021_survey_of_problematic_database_code_fragments}, SQL code smells~\cite{The_119_SQL_Code_Smells, Sharma_2018_Measuring_and_Understanding_Database_Schema_Quality, Muse_2020_Prevalence_Impact_and_Evolution_of_SQL_Code_Smells}, and performance issues~\cite{Chen2014_Detecting_Performance_Anti_Patterns_for_ORM, Yan_2017_Understanding_Database_Performance_Inefficiencies_in_Real_World_Web_Applications, Yang_2018_Structure_Database_Backed_Web_Applications, Shao_2020_Database_Access_Performance_Antipatterns}. Specifically, \citet{Brass_2004_Semantic_errors_in_SQL_queries} proposed a list of semantic errors in SQL queries. 
The book by~\citet{Karwin_2010_SQL_Antipatterns_Avoiding_the_Pitfalls_of_Database_Programming} provides an overview of SQL design anti-patterns containing four categories: logical database design anti-patterns, 
physical database design anti-patterns, query anti-patterns, and application development anti-patterns when employing SQL in the application code. A recent survey by~\citet{Alshemaimri_2021_survey_of_problematic_database_code_fragments} summarized categories of SQL anti-patterns and framework-specific (e.g., ORM) anti-patterns. \citet{The_119_SQL_Code_Smells} documented 119 SQL code smells, concerning database design issues, table design, data types, expressions, naming, routines, query syntax, and security loopholes. \citet{Shao_2020_Database_Access_Performance_Antipatterns} conducted a literature survey and reported 34 database access performance anti-patterns in total. 

Despite these efforts, there is a lack of study toward understanding database access bugs using SQL queries or ORM frameworks. Database access bugs in database-backed applications during runtime may be different from the syntactic or semantic errors in separate SQL queries since database access leverages SQL queries embedded within the application code or generated by the ORM frameworks to interact with DBMSs. On the other hand, unlike SQL anti-patterns or SQL code smells, which allow programs to execute correctly but have quality problems such as poor performance or indicate the presence of quality problems but not necessarily bugs, respectively, database access bugs may cause severe problems like crashes. Our work addresses this gap by providing categories of database access bugs from the issue tracking system and highlighting their root causes. We believe that the category and our actionable implications would help developers understand the maintenance issues and challenges in database-backed applications.


\phead{Adequacy of tests in database-backed applications.}
Prior works have proposed different coverage criteria to measure the adequacy of tests for database-backed applications, e.g., SQL commands~\cite{Halfond_2006_Command_Form_Coverage_for_Testing_Database_Applications}, SQL queries and clauses~\cite{Tuya_2004_Using_SQL_Coverage_Measurement_for_Testing_Database_Applications, Tuya_2010_Full_Predicate_Coverage_for_Testing_SQL_Database_Queries}, and database scheme constraints~\cite{Mcminn_2015_Test_Coverage_Criteria_for_Relational_Database_Schema_Integrity_Constraints}. \citet{Halfond_2006_Command_Form_Coverage_for_Testing_Database_Applications} introduced a database interaction testing adequacy criteria based on \textit{command-form} coverage which takes into account variants of SQL commands. \citet{Tuya_2004_Using_SQL_Coverage_Measurement_for_Testing_Database_Applications} introduced a coverage metric for \texttt{SELECT} queries while \citet{Tuya_2010_Full_Predicate_Coverage_for_Testing_SQL_Database_Queries} proposed a form of predicate coverage criterion for SQL queries by considering the coverage of several clauses like \texttt{JOIN}, \texttt{WHERE}, \texttt{HAVING}, \texttt{GROUP BY}, etc. \citet{Mcminn_2015_Test_Coverage_Criteria_for_Relational_Database_Schema_Integrity_Constraints} proposed a family of coverage criteria for testing the integrity constraints in a relational database schema. However, these coverage criteria are defined separately for embedded SQL queries. Our findings show that database access bugs may also be related to database access API calls, configuration, or converting SQL query results, suggesting the introduction of corresponding coverage criteria for tests in database-backed applications.

Other works have applied mutation testing for SQL queries to assess the adequacy of tests in database-backed applications and proposed mutation operators for SQL queries~\cite{Yue_2011_Survey_of_Development_of_Mutation_Testing, sarkar_2013_testing_db_apps_using_coverage_mutation_analysis}. \citet{Chan_2005_Fault_based_testing_of_database_application_programs} proposed seven SQL mutation operators based on the enhanced entity-relationship model. \citet{Tuya_2006_SQLMutation, Tuya_2017_mutating_database_queries} proposed a set of mutation operators for SQL queries and integrated these operators into a tool called SQLMutation that automatically generates mutants for SQL queries. \citet{Zhou_2011_JDAMA_Java_Database_Application_Mutation_Analyser} extended the work by Tuya~\etal and applied mutation testing to database application programs by performing those mutation operators on the SQL queries in Java/JDBC applications. \citet{Gupta_2010_X_data_Generating_test_data_for_killing_SQL_mutants} proposed a set of join/outer-join mutations that model common programmer errors. \citet{Kapfhammer_2013_Search_Based_Testing_of_Relational_Schema_Integrity_Constraints} and \citet{Wright_2014_Equivalent_Mutants_on_Database_Schema_Mutation_Analysis} also proposed operators for introducing SQL query faults that violate database schema constraints. However, all the above mutation operators for SQL queries are proposed aiming at covering SQL features or SQL syntax and semantics without considering real SQL query bugs. 
Our study of real database access bugs (especially the SQL queries and database schema-related bugs) from open-source applications, ranging from 5 to 16 years, can be used as a complementary aid for designing mutation operators and therefore effective SQL mutants~\cite{McMinn_2019_Ineffective_Mutants_for_the_Mutation_Analysis_of_Relational_Database_Schemas}. Besides, our findings show that quite a number of database access bugs are introduced when calling database access APIs (21.7\%) or converting SQL query results (6.8\%), which calls for the study of mutation for host languages (e.g., Java) that access databases by issuing SQL queries.

\phead{Database schema and program co-evolution.} 
Prior research~\cite{Qiu_2013_Empirical_Analysis_of_Co_Evolution_of_Schema_and_Code_in_Database_Applications, Chen_2016_An_Empirical_Study_on_the_Practice_of_Maintaining_Object_Relational_Mapping_Code_in_Java_System,curino08schema, Maule_2008_Impact_Analysis_of_Database_Schema_Changes} study the co-evolution between database schemas and database-backed application programs. \citet{curino08schema} studied the database schema evolution on Wikipedia and the effect of schema evolution on the system front-end. \citet{Maule_2008_Impact_Analysis_of_Database_Schema_Changes} proposed a program analysis-based approach to perform change impact analysis on applications caused by database schema changes. \citet{Qiu_2013_Empirical_Analysis_of_Co_Evolution_of_Schema_and_Code_in_Database_Applications} conducted a comprehensive empirical analysis of how programs co-evolve with schema changes in database-backed applications.
They find that database schemas evolve frequently and schema changes induce significant code-level modifications. \citet{Chen_2016_An_Empirical_Study_on_the_Practice_of_Maintaining_Object_Relational_Mapping_Code_in_Java_System} reported that in particular ORM-related code, changes are more scattered and frequent than regular code and these changes mostly address performance or security concerns. Previous studies mainly focus on how database schema changes impact applications programs, while we investigate how they may cause bugs in applications. In our work, we generalize categories of the root causes of database access bugs, of which some bugs are caused by database schema changes. Our derived categories provide a finer-grained view of the maintenance challenges that developers encounter.

%% file: texFiles/conclusion.tex
\section{Conclusion}\label{sec:conclusion}
In this paper, we conducted an empirical study of 423 database access bugs from seven real-world Java database-backed applications. 
We find that although the number of reported database and non-database access bugs share a similar trend, their modified files in bug fixing commits are different. This implies that they are not necessarily co-located and database access bugs may have different causes as compared to non-database access bugs.
Execution of SQL queries forms an integral part of database accesses and weighs 40\% (169/423) of the studied bugs. The downside is that errors in SQL queries are not checked at compile time, thereby leading to erroneous access to the database. Further, many problematic SQL queries are not fully covered by test cases. This suggests that future development of test generators for database-backed applications should target SQL query coverage. Additionally, databases are the key components of database-backed applications and bugs related to database schema weigh 22.5\% (95/423) of the studied bugs. This calls for the development of test cases that comprehensively cover the database (e.g., database constraint).
We believe that our findings would provide developers with a comprehensive understanding of database access and aid related research on bug detection, testing, and debugging regarding database access.